# Order induces toughness in anisotropic colloidal crystal composites

Victoria Vilchez[1]†, Shitong Zhou[1]†, Florian Bouville[1]*

[1]Centre for Advanced Structural Ceramics, Department of Materials, Imperial College London, London (UK)
†These authors contributed equally to this work.
*Corresponding author. Email: f.bouville@imperial.ac.uk

Spatial ordering of matter elicits exotic properties sometimes absent from a material's constituents. A few highly mineralized natural materials achieve high toughness through delocalized damage, whereas synthetic particulate composites must trade toughness for mineral content. We test whether ordering the mineral phase in particulate composites through the formation of macroscopic colloidal crystals can trigger the same damage resistance found in natural materials. Our macroscopic silica rod based anisotropic colloidal crystal composites are processed fully at room temperature and pressure, reach volume fraction of mineral higher than 80%, and aided by a ductile interface, unveil plastic strain reaching 10% through the collective movement of rods and damage delocalization over millimeters. These composites demonstrate key design rules to break free from conventionally accepted structural materials properties trade-off.

Spatially ordering a material microstructure can make properties absent from its composition emerge, giving rise to the concept of metamaterials. These unusual behaviors stem from a spatial variation of a properties leading to structural color [1], negative refractive index [2], or acoustic bandgaps [3]. Whereas working on the effect of structure and order [4–9] in porous lattice-based materials led to numerous breakthroughs in mechanical properties, the same cannot be said about dense materials. First, it is problematic to fabricate dense and ordered microstructures at the macroscopic scale and it is difficult to predict how mechanical properties would be affected. Toughness, a material's resistance to fracture, is a property that depends on both the microstructure and the constituent properties [10]. It is linked to a material capacity to plastically deform and its long-term durability and fatigue resistance. More importantly, a high toughness is responsible for users' security in safety critical applications, such as aeronautics or nuclear reactors [11,12], but also dictates the performance of functional materials necessary for the energy transition [13]. Numerous microstructural changes have been tested to improve materials' toughness and deformability by relying on mechanisms acting at the atomic [14,15] or molecular length scale [16]. Introducing order in the microstructure could provide a universal and potent way of spreading damage and delaying failure by erasing the presence of a weak path for fracture, breaking us free from the trade-off observed in structural materials [17].

Examples of such intricate control of the microstructure can be seen in natural materials and have fascinated researchers because of their combination of properties absent in man-made materials [18]. In nacre's brick-and-mortar structure, toughness amplification is due to the regularity in brick-and-mortar dimensions combined with the strain hardening of the interface [19–22]. Simulation work has even proved that altering even slightly this ordering of the microstructure fatally removes the damage tolerance [23]. The dactyl club of the mantis shrimp [24,25] or tooth's enamel [26] are also associated with high hardness, damage and wear resistance through mineral rods and organic mortar arrangement. The common denominator in all these structures is the ability of natural materials to control the local packing and to order an anisotropic mineral phase and a ductile phase over large distance [27]. Plenty of materials got inspired by these feats, leading to composites with impressive properties and microstructures featuring reinforcements in a wide range of sizes, from the 100 μm down to the 100 nm scale [18,28–31]. However, none of the materials made so far present an ordering of the microstructure's constituents as good as what has been found in natural materials and thus none presented the extent of damage delocalization and toughness amplification seen in natural materials. The only examples that managed to mimic this ordering were done with macroscopic reinforcing elements, proving again that order is key to observe damage delocalization [32,33]. More generally, it is a long-standing trade-off observed in any synthetic composites: the higher the volume fraction of reinforcement in a ductile matrix, the lower the toughness and plastic deformation [34,35].

The only synthetic materials that come close or even exceed the order found in natural materials are colloidal crystals. These crystals made from monodisperse particles of size ranging from a few nanometers to a micron can self-assemble into larger structures [36], forming both useful functional materials and models to study atomic structures [37]. While most colloidal crystals are made of spheres, recent breakthrough in sol-gel synthesis led to monodisperse rods of silica in the micron range [38]. These rods can assemble into structures similar to liquid crystals, their molecular analogue, and are already opening new ways to look at phase behavior and crystal defects in anisotropic crystals [39,40]. As most toughening mechanisms in composites rely on anisotropic elements, having anisotropic mineral particles that can assemble into colloidal crystals finally opens up a whole new field of study on the effect of order on damage resistance. There is no intrinsic limit to the volume they could reach even if the number of defects will increase with the crystal size [41]. However, today most colloidal crystals are limited to a few hundreds of microns in size due to the difficulty in controlling their entropy-driven assembly and growth [42–49].

The objective of this study is thus two-fold: to fabricate centimeter-sized colloidal crystals with anisotropic particles so they can be mechanically tested, and to study the effect of ordering on the damage delocalization in composites.

### Fabrication of macroscopic anisotropic colloidal crystal

Fig. 1 describes the process of templating the entropy-driven self-assembly of monodisperse rods to obtain centimeter-sized anisotropic colloidal crystal composites (a-$C^3$) at close to room temperature.
Starting from the room temperature sol-gel synthesis of silica rods developed by Kuijk *et al.* [38], we developed a templated method to grow large scale colloidal crystals (Fig. 1a-c). The rods, 3 µm in length and 300 nm in diameter (Fig. S1), are first functionalized using $\gamma$-MPS to facilitate the future infiltration of acrylate monomers and strengthen the interface between the rods and the polymer interface (Fig. 1a). A template is introduced at the bottom of the mold during the crystallization process to guide the orientation of the crystals growing and obtain a textured polycrystal of large dimension [50]. The template is based on an off-the-shelf optical grating that is imprinted on a silicone substrate using a soft-lithography method. The template consists of periodic wedges angled at 33° from the horizontal and separated by 1.67 µm (Fig. S2). The rods are mixed in an index matching solvent made from DMSO and water to decrease the magnitude of attractive interparticle forces, enabling crystallization during sedimentation. A rod slotting in the wedge will lose 0.4 $k_BT$ of gravitational energy, where $k_B$ is the Boltzmann constant and T the temperature, providing a driving force to control the orientation of the rods in the crystals growing from the template (Fig. 1b). The two conditions to ensure that the crystallization of the rods starts from the template and can grow from it are: to have an initial volume fraction of rods well below the isotropic-to-nematic phase transition, and that the magnitude of the driving force for crystallization is comparable with the rod's Brownian diffusion. The Péclet number in this condition is $P_e = 1.22$ (see Methods), confirming a similar contribution between advection from gravity and Brownian motion in the movement of the rods. The final step of the fabrication consists in drying the crystal formed before infiltrating it with low viscosity acrylate monomers. The composites are then polymerized under UV light, forming large scale anisotropic colloidal crystal composites (a-$C^3$).

We first confirmed the formation and size of the rods from the sol-gel synthesis and their capacity to self-assemble upon sedimenting using SEM (Fig. 1d-e). Having the rods dispersed in an index matching solvent allows to have a fully transparent solution when the rods are in the isotropic phase, which becomes translucent as the rod concentration increases. We used an optical setup with cross polarizers to follow the growth of the crystal over time with or without the template (Fig. 1f). The crystal growth front is visible in both configurations, however the sample without template presents a visibly more heterogeneous structure, with multiple grains of different colors present after 164h. This suggests that multiple crystals with different orientations are present in the non-templated conditions, each crystals changing the light polarization differently. The position of the interface as a function of time provides quantitative information on the crystal growth. The templated growth presents a faster initial growth compared with the non-templated one, with a crystal thickness 3 times larger after 20h. However, the growth speed appears similar after this initial burst. The faster initial growth in presence of the template points toward a faster nucleation of the crystal with the template, an effect also seen in other colloidal crystal systems and thought to occur in seeded crystal growth of any kind [51].

Finally, we demonstrate that crystals of several millimeters in height can be obtained from a templated growth after several days and that their optical properties suggest a uniform orientation of the rods throughout the whole sample. This process is *a priori* scalable to larger sample size by enlarging the template area and longer sedimentation time. While this growth time can seem long, we manage to obtain composites in the $mm^3$ range in two weeks, from rod synthesis to composite fabrication. It seems a necessary price to pay to obtain the high degree order in the microstructure and to keep the embedded-energy necessary to fabricate these a-$C^3$ as low as possible.

We can *a priori* grow cm-scale colloidal crystals from this simple process at close to room temperature, but the local structure and orientational control over macroscopic distance is key to study the mechanical properties.

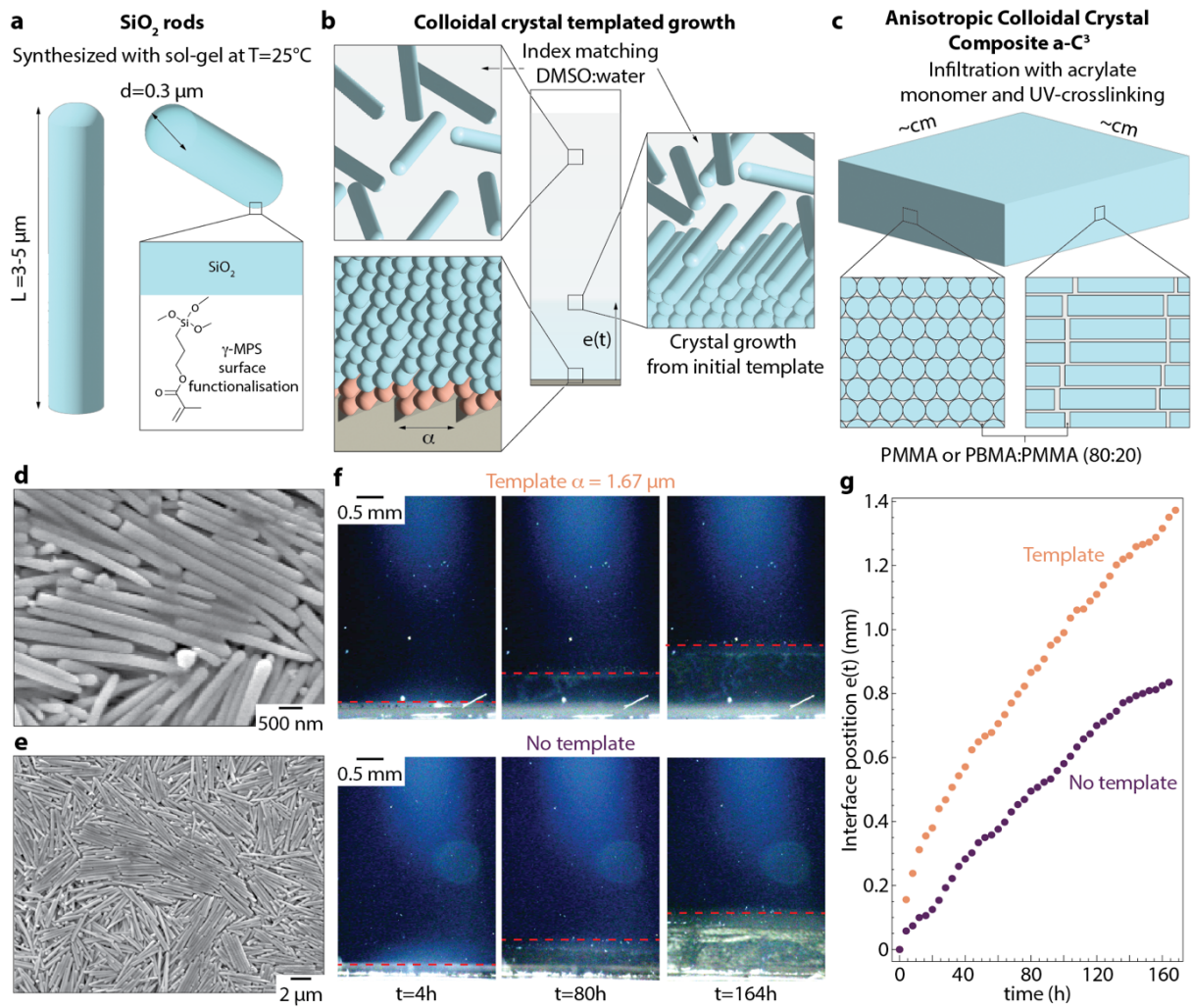

Fig. 1. Fabrication of bulk ordered colloidal crystals from anisotropic building blocks. Schematic representation of the composite fabrication: (a) sol-gel synthesis of rod and functionalization with $\gamma$-MPS, (b) templated entropy-driven assembly of the rods in DMSO-water mixture into cm-sized colloidal crystal, (c) infiltration with acrylate monomer after solvent-removal and cross-linking to form the anisotropic colloidal crystal composite (a-C$^3$). (d, e) SEM images of the as-synthesized rods. (f) Optical tracking of the crystal growth with and without template through a cross-polarized microscopy setup. (g) Position of the disorder-to-crystal interface as a function of time.

### Characterization of the short- and long-range order in the composites

We confirmed using small-angle X-ray scattering (SAXS) and electronic microscopy that we can control the orientation of rods over the millimeter scale and thus grow some of the largest textured colloidal polycrystals made with a local packing close to the theoretical limit.

SEM images of the top surface of composites grown with and without template show that the rods are indeed self-assembled into a smectic phase in the composite, with layers of rods stacked into columns with their ends almost aligned. While the sample grown with a template shows an alignment of rods over tens to hundreds of microns (Fig. 2a), the sample grown without a template displays changes in orientation of the rods over distances around tens of microns. Multiple topological defects expected from free growth of anisotropic colloidal crystals are present in the non-templated sample, with for instance a disclination line visible in Fig. 2b. Even in the templated sample the rods main orientation changes over long distances, however, the orientation changes around a common direction. This observation is coherent with what we expect from the templated growth, with multiple crystals growing at the same time from the template with a common rod orientation.

Using ion-polishing, the smectic packing of the rods becomes even more apparent, while the polymer layer with a thickness of around 30 nm can also be more easily seen (Fig. 2c). A fast Fourier transform (FFT) of the image displays a cross-shaped feature, revealing that the stacking of rods in the image displays a preferred orientation, but also that the ends of the rods form an additional pattern oriented at 90° from the rod stacking direction. An SEM image of the ion-polished cross-section taken perpendicular to the template direction provides further insight into the packing of the rods (Fig. 2d). The FFT of the whole image reveals a periodicity in the distance between the rods, with a ring visible in it. Rods can even form a more ordered structure locally, with an FFT highlighting the presence of local hcp packing. The packing of rods was quantified using image analysis of multiple images and reaches 82%±2% and 64%±15% for the templated and non-templated composites respectively. The templated crystal packing values are closed to the theoretical limit for hard rods of 91 vol%, which is higher than most composites fabricated around room temperature where the packing is limited by reinforcement polydispersity, and as high as some composites made using multiple pressing and heating steps [52,53].

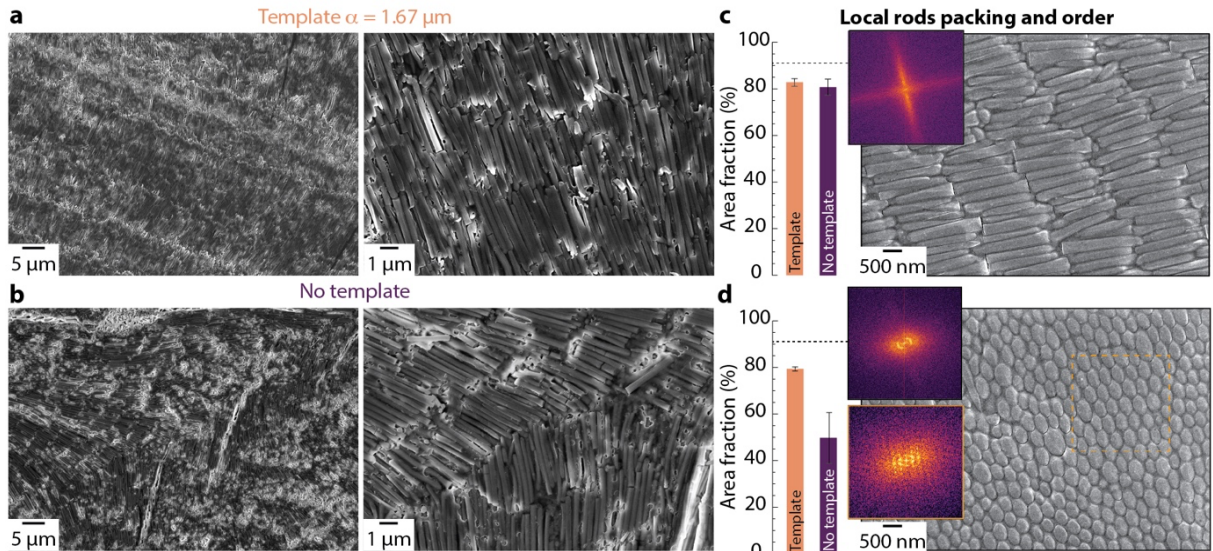

Fig. 2. Short-range order of the rods in a-$C^3$. SEM images of the structure of composites fabricated (a) with a template and (b) without. SEM images of ion-polished cross-section of the a-$C^3$ grown from a template with cross-section taken (c) in the wedge direction and (d) perpendicular to the wedge direction. Insets are FFT of the images or of highlighted area in the image. The dotted line at 91% in the histograms is the theoretical maximum volume fraction of rods.

While SEM provides evidence of both the short- and long-range control over the rod packing and orientation provided by the templated growth method, obtaining samples with this ordered structure over millimeter to centimeter size is necessary to measure the effect of order on the mechanical properties of a-$C^3$. We decided to use SAXS to characterize the orientation and order over larger volumes. The SAXS beam probes a disk of 200 µm in diameter through the whole sample thickness, and we recorded the SAXS pattern at 9 different spots spaced by 1.5 mm, covering a total zone of 3 x 3 mm² (Fig. 3a). The isolines at 50% of scattered intensity at the 9 positions for a sample grown with and without template with the X-rays taken along the growth direction are represented in Fig 3a. All the patterns obtained in the templated sample present a preferred orientation at 90° from the direction of the template, whereas the sample without template display a more isotropic response. The SAXS patterns are reminiscent of the FFT obtained in Fig. 2c (and Fig. S3), with a cross shape visible although less clearly, indicating that the rods columns are less ordered in the bulk than locally. The azimuthal profiles confirm the common orientation (Fig. 3b), with a major peak visible around $\phi$=150° in all patterns. The position of this primary peak across the 9 scans indicates an alignment within ±8° of the rods' orientation around the template direction. These results prove that our a-$C^3$ composites present a common orientation over several millimeters, *i.e.* over scales 4 orders of magnitude higher than the average rod length. The radial profiles along the primary direction, at 90° from the template direction, and the secondary direction, at 0° from the template direction, do not show any features (Fig. 3c), proving that there is no periodicity of the rods' arrangement over long distance. However, the cross-sectional radial profiles taken when the X-rays traverse the sample in the direction of the template orientation show rings at a wave vector of 0.0042 Å$^{-1}$ (Fig. 3d). This wave vector corresponds to around 150 nm in real space. We attribute this ring to the second order diffraction generated by the periodicity in rod-to-rod distance corresponding to their 300 nm diameter.

The study of the microstructure of the a-$C^3$ proves that the templating method successfully orients the rods over areas of several millimeters squared and thicknesses of several millimeters. These bulk anisotropic textured colloidal polycrystals are some of the largest reported, with thicknesses one order of magnitude higher than other templating methods used with isotropic colloids so far (Fig. 3e). The testimony of this control of the rod orientation over macroscopic distances can also be seen in their optical properties, with a visible shimmering (Fig S4) and even the presence of structural colors when observed under a microscope (Fig. 3e inset) when no polymer is present between the rods.

We can now fabricate millimeter scale colloidal textured polycrystals, but the final mechanical properties will be dictated by the local structure and the rods and polymer properties.

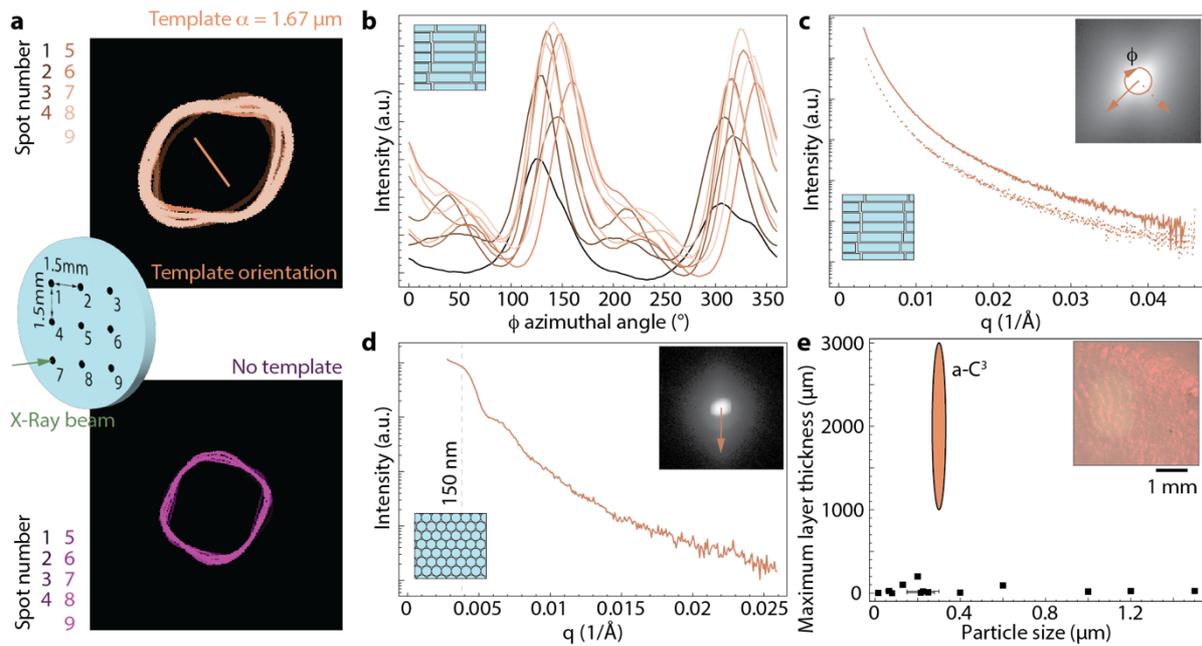

Fig 3. Long-range order of the rods in a-C³. (a) Line representing the intensity at half maximum of the SAXS signal made at 9 different spots on sample grown with and without template with the X-rays traversing the sample in the direction of the crystal growth. The line in the top image represents the orientation of the wedge of the template. (b) Integrated intensity as a function of the azimuthal angle $\phi$ for all scans taken on the sample grown with a template. (c) Intensity as a function of wavenumber in the direction of the template and perpendicular to it. (d) Intensity as a function of wavenumber with the X-ray going along the direction of the template. (e) Comparison of the size of the crystal grown with the silica rods in this work with other methods to grow large size colloidal crystal with spheres. Inset: optical microscope image of an a-C³ taken in the direction of the sample growth before infiltration. Data from references [42–49,54–59].

### Damage delocalization in the anisotropic colloidal crystal composites

Now that we can obtain macroscopic highly mineralized composites with short- and long-range order of reinforcement, we are finally able to test how this microstructure responds to mechanical loads and damage. Taking a page out of natural materials' book, we know that the properties of the interface are instrumental in enabling damage tolerance. We test this hypothesis by using two different acrylate based-polymer with vastly different properties as interface: PMMA, a commonly used linear elastic brittle polymer [60–62], and a mixture of PBMA:PMMA that presents a plastic strain of 200% at the expense of rigidity and strength (Fig. 4a). We performed flexural tests *in situ* in an SEM to visualize the damage in the highly mineralized composites with a fine spatial resolution.

While the presence of PMMA at the interface leads to an almost brittle behavior of the composites (Fig. 4a and Fig. S5), the introduction of a deformable interface enables plastic strain as high as 10% in this 80 vol% mineral composite. This high strain value is even more impressive if we compare it with the ones obtained in particulate composites: regardless of the composition, size, and shape of the ceramic reinforcements, as well as strain at failure of the interface, particulate composites present strains at failure within 0 to up to 2% in the 60-98 vol% of reinforcement range (Fig. S6). We attribute this 5 to 10-fold increase in strain at failure to the order introduced in the reinforcement through colloidal crystal assembly and the high strain at failure of the polymer at the interface. The large strain obtained is macroscopically visible as our samples are significantly bent after testing (Fig. 4c).

To confirm the role of the rods' ordered packing in the large deformation at failure of the composites, we observed the damage on the face in tension during flexural testing using electron microscopy *in situ* (Supplementary Video 1). Only a few cracks are visible at the composite yield point around 1% strain (Fig. 4a, b), but they quickly multiply once the macroscopic strain reaches 4%. The composite at 7% strain presents a dense network of cracks, with a spacing on the order of the rods' length. We validate this with an FFT of the crack network obtained at 7% strain, and the image confirms that the crack network is oriented vertically and presents some periodicity that falls in the range of the rod's length (Fig. 4d). The network of cracks follows the smectic arrangements of the rods' assembly, leading to a multitude of fine cracks spread out in the microstructure instead of a single crack that would concentrate the stress locally and lead to the failure of the part.

This ordered architecture thus seems powerful at avoiding the creation of a major crack, a capacity that can be further probed during fracture testing. The resistance to crack propagation of the a-$C^3$ | PBMA:PMMA was quantified using conventional macroscopic single edge notch bending (SENB) test and electron microscopy to follow the damage in real time. The toughness of the composites as function of crack extension, also called R-curves, are plotted in Fig. 4e along with the toughness of the PBMA:PMMA polymer used at the interface. All the composites present a rising R-curve, with the toughness increasing over the first 70 µm by a factor 4 to 8-fold, compared with the polymer (Fig. S7). This extent of toughness amplification compared with the interface material is visible only in a few natural materials, mainly nacre, and is associated with the collective sliding of bricks over millimeters. Crack deflection, twisting, or pull-out toughening mechanisms could not lead to such a high toughness amplification (see supplementary discussion). In addition, the rapid increase in toughness with the initial crack extension can be linked with the stability of the fracture process: the higher the toughness, the higher the necessary stress to reach an unstable fracture [63].

The toughness increase is directly linked with the formation of a damage zone in front of the crack full of rod columns sliding (Fig. 4f). This damage zone increases in size in front of the crack and prevents any further propagation of the major defect, proving the capability for damage delocalization of the a-$C^3$ structure. Even after the damage zone is fully developed and the main crack starts to extend, its propagation is surrounded by a process zone reaching 0.5 mm, thus making a hundred rods within one line move (Fig. 4f). Approximating the process zone as a cylinder with a diameter of 0.5 mm and a depth equal to the depth of the sample (Fig. 4g), we calculate that up to around $5 \cdot 10^8$ rods are collectively moving within the process zone to prevent the crack from growing.

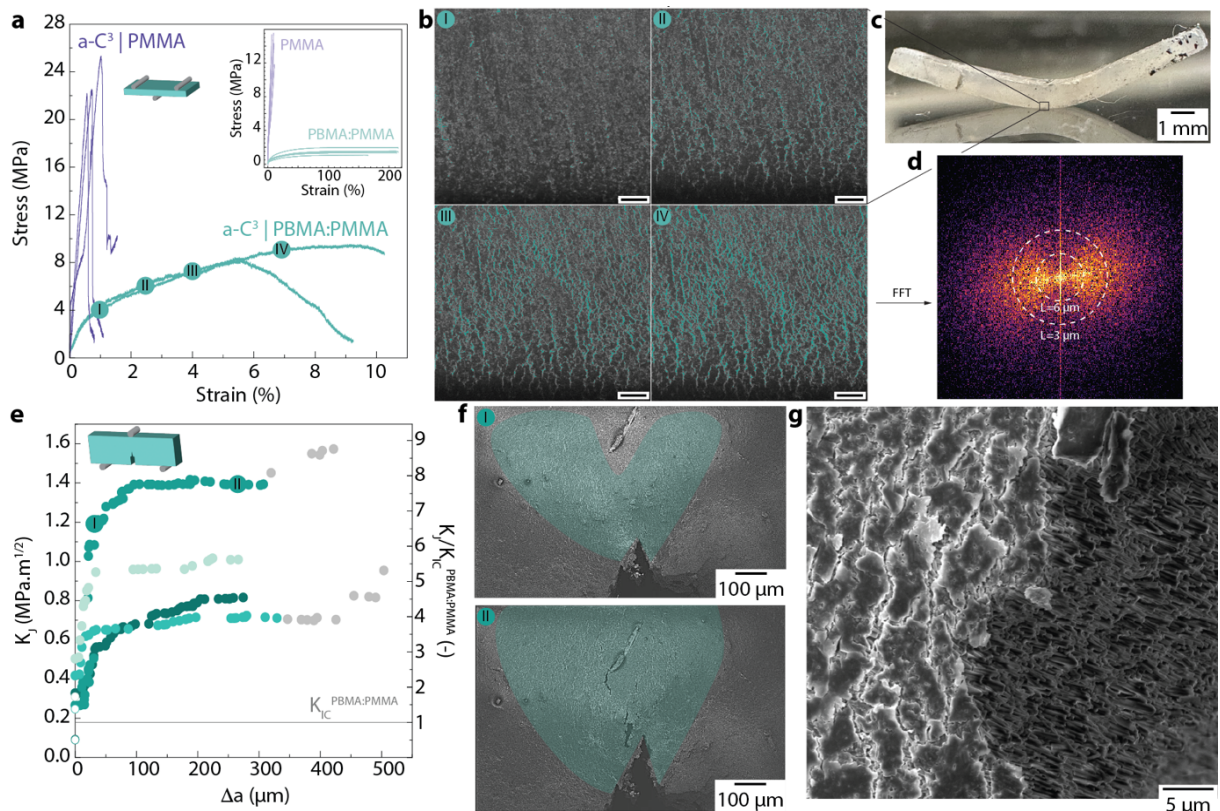

Fig. 4. Mechanical macroscopic behaviour of anisotropic colloidal crystal composites a-C[3]. (a) Stress-strain curves for a-C[3] composites tested in bending with two polymer interfaces. Inset: Stress-strain curves of the pure polymers used as interfaces in the composites tested in tension. (b) SEM images of the zone in tension taken at different strains during *in situ* bending test of a-C[3] | PBMA:PMMA. Cracks are highlighted in teal. Scale bar 20 µm. (c) Image of the a-C[3] | PBMA:PMMA after bending test. (d) FFT of the cracks network visible in the bottom part of the sample. (e) R-curve measured from Single Edge Notch Bending tests of a-C[3] | PBMA:PMMA. Empty symbols represent measurements in which microcracks are present in front of the main crack before it propagates. Grayed symbols represent values beyond the ASTM-recommend crack extension limit. (f) SEM of the crack front taken at different crack lengths during the *in situ* fracture test. Microcracked area highlighted in teal. (g) SEM of the region close to the main crack.

## Conclusion

In conclusion, we introduced order into particulate anisotropic composites using a combination of principles from colloidal crystal, traditional crystal growth, liquid crystal, and composite fabrication. Because of that, our anisotropic colloidal crystal composites are manufactured entirely at close to room temperature and pressure, reach macroscopic sizes, and still present up to 80 vol% of mineral reinforcements. The monodispersity of the rods and the entropy-driven templated assembly leads to a smectic microstructure that is present at short range but also at long range as confirmed by a combination of microscopy and diffraction results. The sol-gel derived rod assembly into highly ordered structure triggers toughening mechanisms at local and macroscopic length scale only seen in natural materials so far. The composites present a macroscopic strain at failure of up to 10%, breaking free of the expected trade-off between high reinforcement content and high deformability and toughness in particulate composites. We established that this capacity to delocalize damage over large volumes is directly rooted in the order of the microstructure and the polymeric interface deformability. The materials produced here are intrinsically interesting to unveil further the influence of microstructure periodicity on mechanical properties at the macroscopic scale. But more importantly, the effect we provide here can be applied to any composition and for any application where high toughness and high mineral content is crucial, from solid electrolyte energy storage devices to high performance structural materials for extreme environments, such as nuclear fusion reactors or space shuttle protection systems.

## Materials and Methods

### Materials

3-(Trimethoxysilyl)propyl methacrylate ($\gamma$-MPS, 98%), ammonia solution (28% in water, >99.99%), 2,2-dimethoxy-2-phenylacetophenone (DMPA, 99%), dimethyl sulfoxide (DMSO), methyl methacrylate, (MMA, 99%), 1-pentanol (ACS reagent, > 99%), polyvinylpyrrolidone (PVP, Mw = 40,000 g/mol), sodium citrate dihydrate (> 99%) and tetraethyl orthosilicate (TEOS, 98%) were purchased from Aldrich-Merck. Ethanol (> 99.5% Ph. Eur., USP) was purchased from VWR. Polydimethylsiloxane (PDMS, SYLGARD™ 184 silicone elastomer kit) was purchased from Dow. n-butyl methacrylate (n-BMA, 99%) was purchased from Thermal Scientific.

### Synthesis of colloidal silica rods

The synthesis is a sol-gel process reported by Kuijk *et al*[38]. 12g of PVP was dissolved in 120 mL of 1-pentanol by stirring for several hours in a 500 mL glass bottle. Then, 12 mL 1-pentanol, 12 mL ethanol, 3.36 mL deionized water and 0.80 mL of a solution of 0.18 M sodium citrate in deionized water were added to the bottle. The bottle was shaken by hand to form the emulsion of water droplets in 1-pentanol stabilized by PVP and sodium citrate. 2.70 mL of ammonia solution was added and the bottle shaken again. Afterwards, 1.20 mL of TEOS was added and the bottle was shaken one last time, before being left still to react overnight. Once all the TEOS had reacted, the synthesized rods underwent a series of centrifuging (5804, Eppendorf) and washing steps under the following cycle: the mixture was first centrifuged at 1500 g for 45 mins before the supernatant was poured out. It was then redispersed in ethanol to be centrifuged at the same speed for 15 mins two times. This step was repeated two more times in water and another time in ethanol. The rods were redispersed in ethanol with three final centrifuging steps at 700 g for 15 mins.

### Rods functionalization

Once the supernatant in the last centrifuging step was removed, the centrifuge tube was placed in an oven at 60°C for at least 15 min to fully evaporate the ethanol. The dry mass of rods was weighed and redispersed with a weight fraction of 10 wt% in a solution of $\gamma$-MPS:ethanol 1:2 in volume. The solution was stirred at room temperature for 24 hours, then centrifuged and washed twice in ethanol with runs at 1500 g for 15 min. Finally, the rods were dried at 60°C for 15 mins.

### Assembly into colloidal crystals

The dry rods were weighed and resuspended in an index matching solution. Index matching was obtained by dispersing rods in dimethyl sulfoxide (DMSO, n = 1.479 at 20°C) then adding deionized water (n = 1.333 at 20°C) dropwise until the sol became transparent. The volume fraction of rods in solvent was chosen to be 2 vol% to ensure the rods are in the isotropic phase before sedimenting. A volume ratio of 10:2.1 DMSO:$H_2O$ had to be used during sedimentation to match the refractive index of functionalized rods, yielding an effective refractive index of n = 1.45. Once the rods had been dispersed in DMSO and $H_2O$, the system was left to self-assemble and sediment for at least 7 days. During the assembly of rod-like colloids, the competition between sedimentation under gravitational forces and Brownian diffusion can be first estimated by the gravitational Péclet number [64]:

$$Pe = \frac{4\pi \Delta \rho g R^4}{3 k_B T}$$

for spherical particles of radius $R$, with $\Delta\rho$ the difference in density between the solvent and particles, $g$ the acceleration of gravity, $k_B$ the Boltzmann constant and $T$ the temperature. By assimilating rods to spheres of radius equal to their radius of gyration

$R = \sqrt{\frac{d^2}{2} + \frac{l^2}{12}}$ = 892 nm [65] with $\Delta\rho = \rho_{DMSO} - \rho_{rods}$ = 1900 − 1100 = 800 kg/m³ and T = 293K, Pe is 1.22.

In order to improve the assembly of crystals in a single orientation on a larger scale, unidirectional templating was used. The templates were made of PDMS by duplicating the pattern on an optical grating with groves spaced by 1.67 µm (ruled diffraction gratings with 600/mm grating, Thorlabs). These PDMS-imprinted stubs were then placed at the bottom of polypropylene syringes (5 mm diameter and 75 mm height, 1 mL Plastipak, BD) for producing crystals in 5 mm diameter disks during sedimentation. The piston and conical tip of the syringe were removed before the stubs were sealed with the syringe using epoxy glue (Araldite Instant 90 s). For upscaling the size of the crystals, glass capillaries (7 mm in width, 14 mm in length and 70 mm in height, CM Scientific) were used as the container for sedimentation. To prevent the crystals from sticking to the wall of the glass capillaries after resin infiltration, fluorinated ethylene propylene film (FEP Release Film Liner, Elegoo) was put on the walls of the glass capillaries.

Composite manufacturing
After 7 days of sedimentation, a sediment of 2-3 mm height had formed at the bottom of the cylinder. The supernatant was removed and the remaining solvent evaporated by placing the sample in a vacuum oven (OV-11, Jeio Tech). To evaporate the water first, the oven was kept at ambient pressure and heated at 70°C for 4 hrs. Then, vacuum was pulled down to -0.1 MPa while keeping the temperature at 70°C to evaporate the DMSO. Once the sediment was dry, it was removed from the oven and infiltrated with the monomer and photoinitiator mix. Two acrylate monomers were tested, methyl methacrylate (MMA) and n-butyl methacrylate (BMA). The photoinitiator was 2,2-dimethoxy-2-phenylacetophenone (DMPA). For infiltration with BMA and/or MMA, the monomers were mixed with 10 wt% DMPA. The monomer mix was then slowly added dropwise from the top side of the sediment, leaving enough time for capillary forces to draw the liquid towards the bottom of the sediment without trapping air bubbles. The infiltration of a liquid with a refractive index close to the one of silica rods led to a change of translucency of the sediment upon impregnation, allowing for a visual confirmation of uniform infiltration of the resin. The infiltrated composite was then consolidated by curing the acrylate resin in a UV chamber (Asiga Flash, Asiga) at 365 nm for 30 mins.

Characterization of crystal growth using optical microscopy under polarized light
Polarized light microscopy was used to monitor the sedimentation of silica rods in the DMSO/$H_2O$ solvent and observe the nucleation of crystalline domains. Imaging the sample under polarized microscopy allows to observe 2 mechanisms in 1 setup. First, the observation of birefringence, characterized by the apparition of 2 refractive indices in a material depending on the light propagation direction. Birefringence can change the polarization of light, meaning that light can still pass through a crossed polarizer-analyzer. In the case of the silica rod composites, birefringence could occur from the optical anisotropy of aligned rods in the nematic and smectic phases [66]. In addition to birefringence, Bragg diffraction can occur due to the periodical slits formed by rods as they assemble into nematic or smectic phases. In the case of silica rods of spacing approximately equal to their diameter d = 300 nm, the colloidal crystal diffracts in the visible light range. Because Bragg diffraction can also modify the polarization of light if the latter is not contained in the plane of diffraction, it results that diffracted light can pass through a crossed polarizer-analyzer too.
To set up the polarized microscopy experiment, glass capillaries of 50 mm in length with a cross section of 0.5 mm x 5 mm and a wall thickness of 0.350 mm were used for rod sedimentation (VitroTubes, VitroCom). The glass capillary was bonded to the PDMS template using a plasma bonding technique. To ensure maximum surface adhesion, the bottom of capillaries was first polished down to 5 µm using SiC paper (CarbiMet S, Buehler). The PDMS stub and polished capillary were placed in the chamber of a plasma cleaner (Femto basic unit type D, Diener Electronic, Germany). An oxygen plasma was generated at a gas flow of 15 sccm and a power of 50 W and held for 30 seconds. The two parts were bonded straight after being taken out of the chamber. The capillary was then filled with the solution of rods at 2 vol% in DMSO and $H_2O$ mixture and placed between a polarized light source (BL-ZW1, Dino-Lite, UK) and an optical microscope with a built-in polarizer (AM7013MZT, Dino-Lite, UK). The polarizer of the light source was oriented at 90° to the polarizer of the microscope in order to achieve complete extinction. Images were captured every 4 hrs during a total time of 7 days.

Characterization of microstructure using scanning electron microscopy
The microstructure of the silica rod composites was first characterized using SEM. Before imaging, the composites were embedded in epoxy (EpoThin 2, Buehler) and polished down to 1 µm using diamond suspensions (DiaPro, Struers). To further flatten the surface and image only the contrast between the silica rods and the acrylate resin, some samples were polished using broad argon ion beam milling (PECS II, Gatan) for 30 mins at 4 keV and 6°. A conductive coating of 10 nm of chromium was sputter coated on the samples prior to imaging (Q150T S, Quorum). Samples were imaged in the SEM at acceleration voltages of 5 kV for the secondary electron detector and 10 kV for the backscattered electron detector (Auriga CrossBeam, Zeiss).

Small Angle X-ray Scattering

Since sending X-rays on a colloidal crystal of lattice spacing of hundreds of nm results in diffraction at very small angles, the silica rod composites were characterized using small angle X-ray scattering (SAXS) at the Diamond Light Source synchrotron facility using the DL-SAXS beamline. Silica rod composites of 5 mm diameter and a few millimeters in height were prepared and cut in slices of 1 mm thickness, with the rods either in-plane or in the cross-sectional view. In addition, a blank sample made of pure acrylate resin was used as a control to ensure that the observed signal was solely due to the interaction of silica rods with the incident beam. Slices were placed in a 6 mm disc solid rack. Scans were taken in ambient conditions, using an Excillium Ga MetalJet source of 9.2 keV and a EIGER2 R 1M detector with a pixel size of 75 μm. Each sample was imaged taking a grid of 3 x 3 measurements per sample, separated by 1.5 mm in each direction. Two sets of measurements were taken with the sample-to-detector distance set to 1 m and 4.5 m. Data processing and visualization was conducted using the DAWN software [67].

Mechanical testing

The strength and toughness of the freshly prepared silica rod composites were determined by *in situ* 3-point bending using a 300 N microtest stage (MT300, Deben UK Ltd), with a span of 11 mm and a test speed of 0.1 mm/min. The composites were tested on the day of the cross linking, we observed a small time dependency of the PBMA:PMMA mechanical properties and thus kept this timing between cross-linking and testing constant. The tensile surface and the surface to be imaged were polished down to 1 μm using diamond suspensions. Three bars of 14×2×1 mm$^3$ for each composition (rods infiltrated with PMMA and PMMA:PBMA=20:80) were test for strength and four bars of 14×3×1 mm$^3$ for toughness. One of the bar for PMMA:PBMA=20:80 was tested with an unloading step after 3% strain to demonstrate that there was plastic strain and is plotted separately in Fig. S8. The stage was placed in a Zeiss Sigma FEG-SEM to monitor crack propagation. The cracks in the SEM images during the test were segmented and reconstructed using interactive top-hat in Avizo 9.3. For the measurement of toughness, single-edge notched beams were pre-notched with a 0.25 mm diamond wafering blade, which was then sharpened manually using a razor blade to obtain a notch length $a_0$ comprised between 0.4W < $a_0$ < 0.5W. For calculation of the toughness, the stress intensity factor $K_i$ was evaluated according to ASTM E1820 [68] with the crack length measured from the SEM image sequences during the test:

$$K_i = \frac{FS}{(BW^{3/2})} \frac{3(\frac{a_i}{W})^{1/2}[1.99 - \frac{a_i}{W} \times (1 - \frac{a_i}{W})(2.15 - 3.93\frac{a_i}{W} + 2.7(\frac{a_i}{W})^2)]}{2(1 + 2\frac{a_i}{W})(1 - \frac{a_i}{W})^{3/2}}$$

where $F$ is the load, $S$ is the supporting span, $B$ is the bar thickness and $W$ is the width, $a_i$ is the crack length.

As a comparison, the strength and toughness of the pure resin (PMMA and PMMA:PBMA) were measured on Zwick/Roell Z010 at a test speed of 1 mm/min. The dog bone samples were machined using a CO$_2$ laser cutter (Omtech SH-G3020 40W) in the dimension described as Type V in ASTM D638 [69]. The toughness of the resin

was evaluated using single edge notch tension test. The stress intensity factor is given by:

$$K_I = \frac{F}{B\sqrt{W}} \sqrt{2 \tan \frac{\pi a}{2W}} \cdot \frac{0.752 + \frac{2.02a}{W} + 0.37(1 - \sin \frac{\pi a}{2W})^3}{\cos \frac{\pi a}{2W}}$$

where $F$ is the applied load, $B$ is the thickness of the dog bone, $a$ is the notch length and $W$ is the width of narrow section in the dog shape.

**Acknowledgments:** The authors would like to acknowledge the Diamond Light Source for access DL-SAXS facility under proposal SM32442. and are grateful to Dr. Sam Burholt for help with the beamtime experiments. The authors would like to acknowledge fruitful discussion on the composite fracture and large-scale rod's alignment with Prof. Finn Giuliani and Prof. Richard Todd and on the rod's mechanical properties with Dr. Oriol Gavalda-Diaz. The authors want to acknowledge Enora Saule, Leonore Giner and Guanghan Zhu for the discussion and work done during their Master's thesis.

**Funding:** This work was supported by the Engineering and Physical Sciences Research Council (grant EP/R513052/1). F.B. and S.Z. acknowledge support from the European Research Council Starting Grant [H2020-ERC-STG grant agreement n°948336] SSTEEL.

**Author contributions:** F.B. conceived the idea, participated to the design of the experiments, supervised the research project, designed the figures, and drafted the manuscript. V.V. designed and conducted the experiments on crystal growth, microscopy, SAXS and initiated the design for mechanical characterization. S.Z. designed and conducted the sample size scale-up, mechanical characterization, and the fracture mechanics analysis. V.V., S.Z., F.B. analyzed the data. All authors contributed to the experiments and discussion of the manuscript.

Supplementary discussions and information

Estimation of the energy lost during rod slotting in the template wedge

A rod of length $L$ and diameter $d$ slotting in a wedge of height $h$ will lose an amount of energy $E_{slot} = V_{rod}\left(\rho_{SiO_2} - \rho_{DMSO}\right) g\, h = \frac{\pi d^2}{4} L \left(\rho_{SiO_2} - \rho_{DMSO}\right) g\, h$, with $\rho_{SiO_2} = 1900\ kg/m^3$ and $\rho_{DMSO} = 1100\ kg/m^3$ the density of silica and DMSO respectively, $g = 9.81\ m.s^{-2}$ the gravitational acceleration. Taking $L = 3\ \mu m$ and $d = 300\ nm$, we obtain that $E_{slot} \approx 0.35\ k_B T$ where $k_B$ is the Boltzmann constant and $T$ the temperature.

Estimation of the contribution of different toughening mechanisms to the toughness of the anisotropic colloidal crystal composites a-C³

Starting from a 2D nematic microstructure as depicted in the first figure of the main manuscript, when the crack encounters a rod it can deflect at most at 90°. We can use Cotterel and Rice [70] approach to kinked crack to estimate the toughness decrease for a deflected crack and thus the additional stress intensity factor needed to keep the crack from propagating.

$$K_I^{defflected} = c_{11}(\theta)K_I + c_{12}(\theta)K_{II}$$

With $c_{11}(\theta)$ and $c_{21}(\theta)$ coefficient that depends on the deflection angle $\theta$. With $\theta = 90°$, the stress intensity from the deflected crack is decreased by almost a factor 2, so deflection could increase the toughness by two with respect to the fracture toughness of the interface in the composite $K^i$. We do not observe a large scale deflection in our composite fracture with the ductile interface PBMA:PMMA so the toughness amplification by deflection is lower than 2.

Deflection alone cannot explain the toughness amplification measured in our composites. Using the work from Barthelat *et al.* [71], we estimate how much the bridging by periodically spaced rods can increase the composite toughness. This estimation of the bridging force and toughness is only valid in a 2D cross section passing by the whole diameter of the rods, the area of the rods being deformed will be smaller in other cross section and thus this is the upper bound of the toughness obtain through bridging. We estimate the closure forces produced by one bridging rod of length $L$ and diameter $d$ and being pull-out by at maximum half of its length from the other side of the crack to be $F = \frac{L}{2} \cdot \tau_i$ with $\tau_i$ the shear strength of the polymer at the interface between the rods. The smectic ordering would lead to no pull-out length as the rods' end would be all aligned within the whole composites. The microstructure and SAXS results show that the rods present various overlap with the adjacent rods layer due to the imperfect smectic packing and non-uniformity of the rods length. We use half of the length of the rods as pull-out length as an exaggeration of the actual pull-out length, so we obtain the maximum bridging toughness amplification. By the same rationale, we estimate that one on two rods are bridging the crack, which is a strong overestimation from what we observed during the *in situ* fracture measurements. The continuous bridging traction homogenized over the bridged crack can be expressed as $t(u) = \frac{F}{2d} = \frac{1}{4}\frac{L}{d}\tau_i$.

The bridging toughness amplification in terms of energy release first $G_b$ can be then obtained using [72]:

$$G_b = 2 \int_0^{u_m/2} t(u) du = \frac{1}{4}\frac{L}{d}\tau_i\, u_m$$

Where $u_m$ is the sliding distance at which the cohesion become zero. We can get a conservative estimate of the interfacial strength as $\tau_i = \sigma_i/2$, with $\sigma_i = 1 MPa$ the strength of the PBMA:PMMA interface. The sliding distance $u_m$ can be estimated from the *in situ* fracture test images to around 0.5 µm, leading to an upper estimation of the bridging toughness of $G_b = 0.75\, J/m^2$. Putting the value in terms of stress intensity factor using $K = \sqrt{G\,E}$ with $G$ energy release rate and $E$ the Young's modulus of the composite we obtain $K_b = 0.05\, MPa.m^{\frac{1}{2}}$. The toughness obtained with the damage delocalization in our composite is thus 16 to 32 times higher than a conservative estimation of the toughness amplification obtained through bridging.

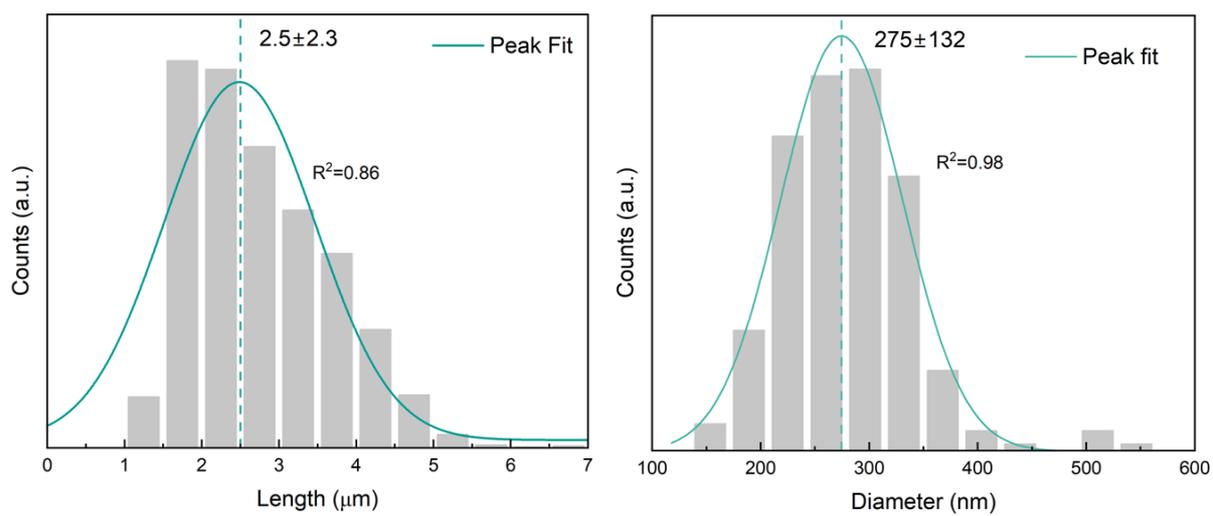

Figure S1. The distribution of rod length and diameter from sol-gel synthesis

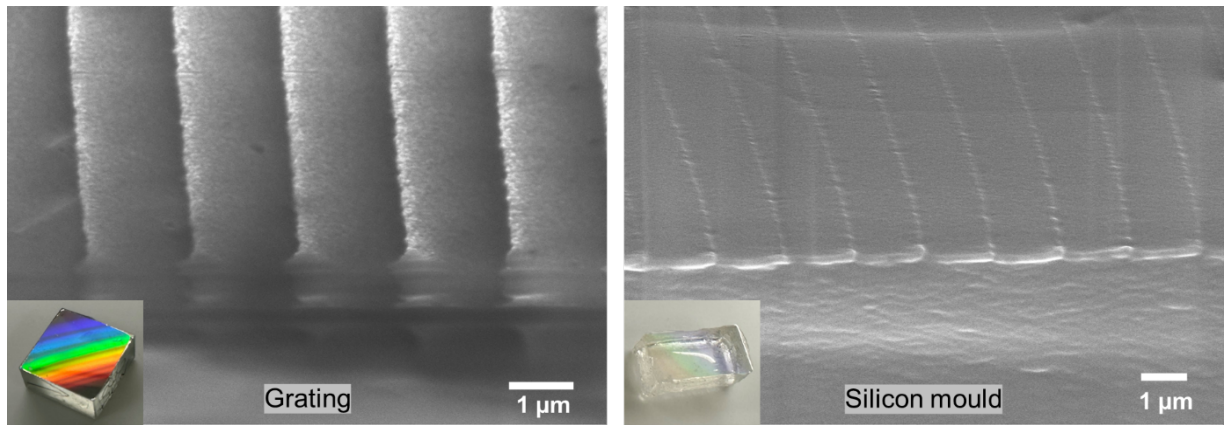
Figure S2. SEM images of the grating and the silicon mould, insert pictures of them showing iridescence

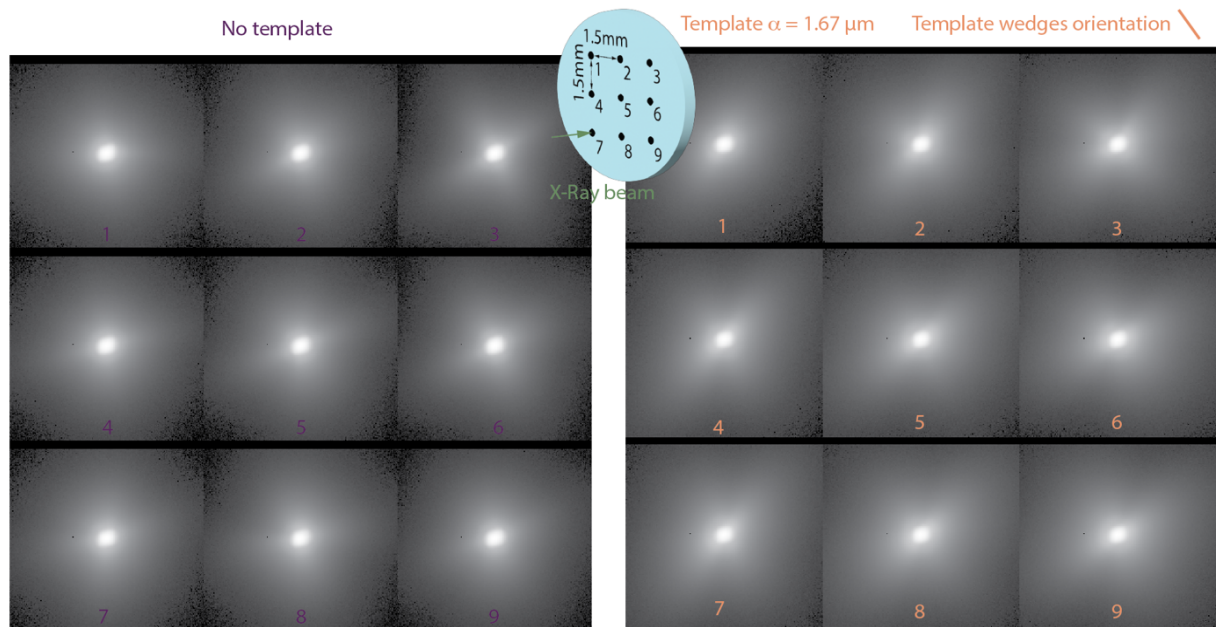

Figure S3. SAXS pattern

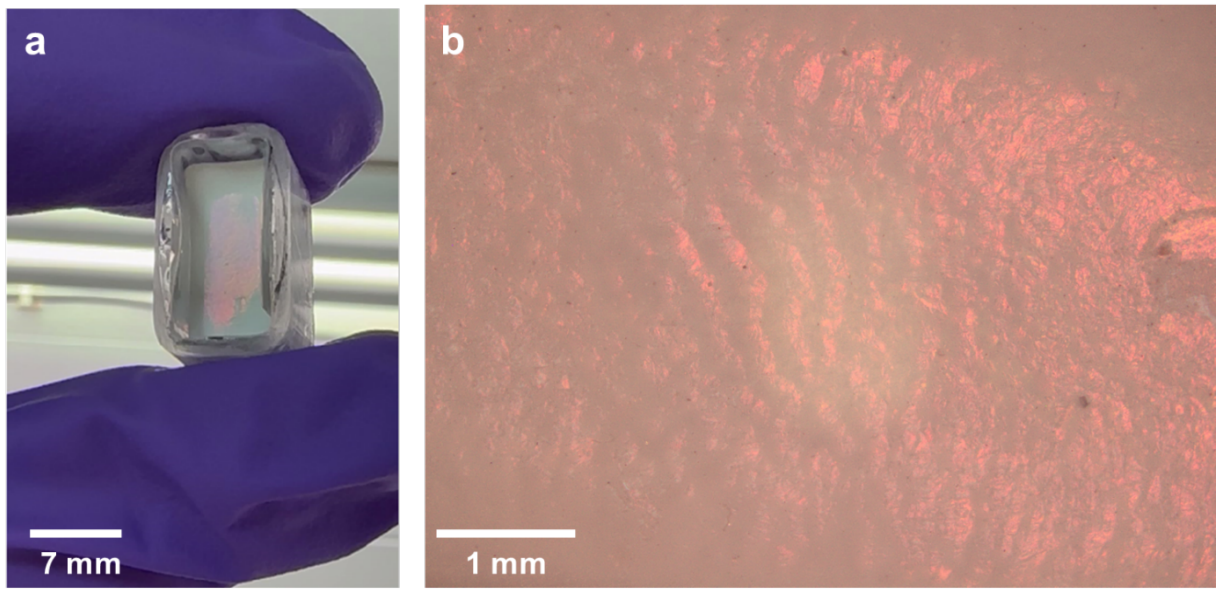

Figure S4. Optical image (a) and microscope image(b) of the dried silica rods, showing visible shimmering and structural colors.

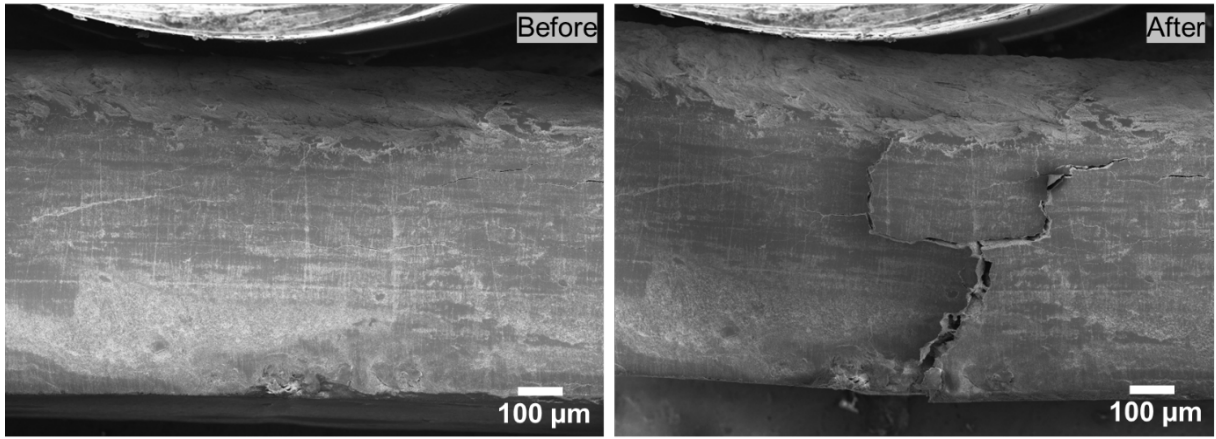

Figure S5. The brittle fracture behaviour of composites with PMMA at the interface

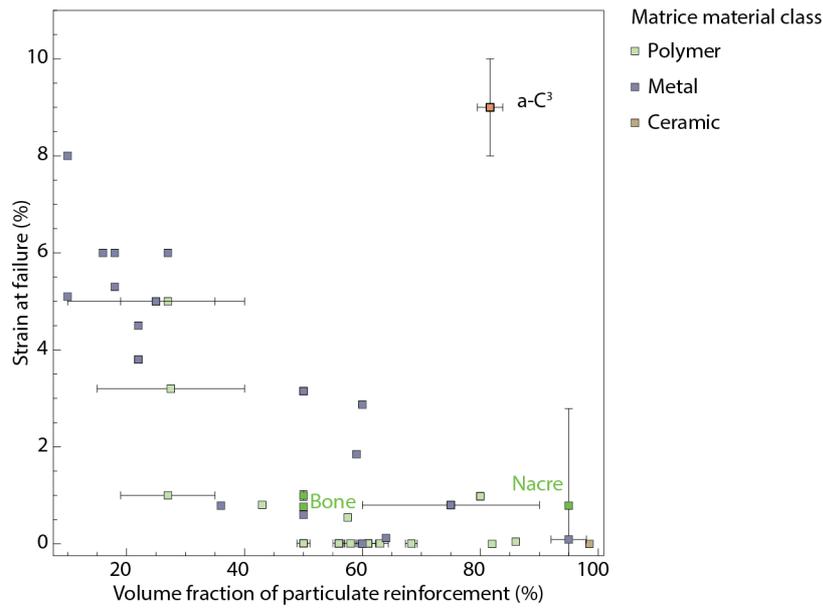

Figure S6. Strain at failure of particulate composites made with different matrices and particulates of different shape but with sizes below 100μm. Data from references: nacre [73], bone [74,75], bioinspired composites with polymer [76–81], metal [82–85], ceramic matrices [86], for conventional composites reinforced by particulate SiC or chopped glass fibres and different matrices are taken from the database of the Ansys CES Granta EduPack software.

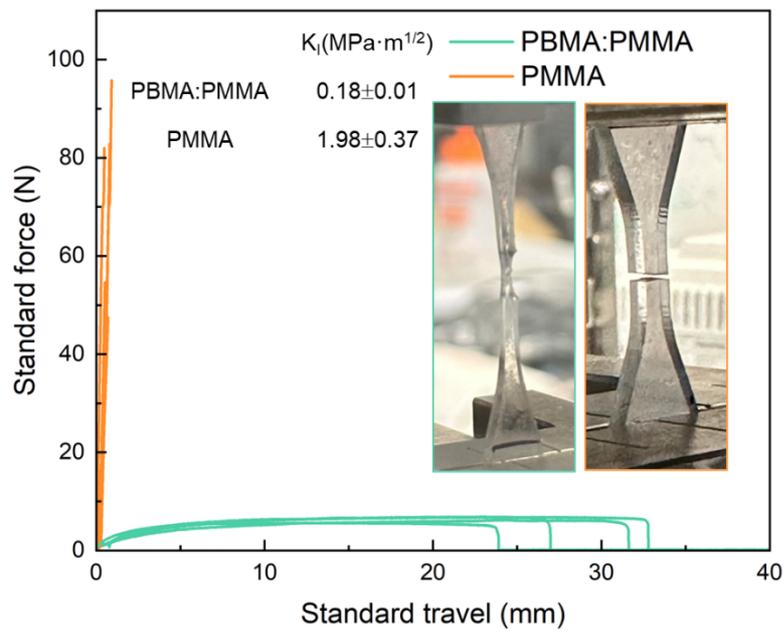

Figure S7. Toughness of the PMMA and PBMA:PMMA from single edge notch tension test.

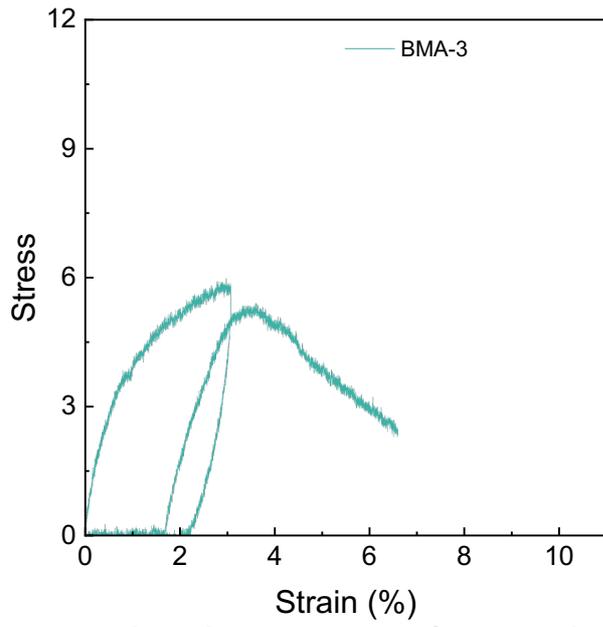
Figure S8. Stress-strain of an a-C3 | PBMA:PMMA in bending with an unloading/loading cycle after 3% strain.